\documentclass[11pt,reqno]{amsart}

\usepackage{latexsym}
\usepackage{a4wide}
\usepackage{amsmath}
\usepackage{amsfonts}
\usepackage{amscd}
\usepackage{amssymb}
\usepackage{amsbsy}
\usepackage{mathrsfs}
\usepackage{yfonts}
\usepackage{bbm}
\usepackage[dvips]{epsfig,graphics}
\usepackage{epsfig}
\usepackage{graphicx}
\usepackage{psfrag}
\usepackage{upgreek}
\usepackage{undertilde}
\usepackage{mathtools}
\usepackage{leftidx}
\numberwithin{equation}{section}
\usepackage{cite}
\usepackage{color}
\usepackage{stmaryrd}
\mathtoolsset{showonlyrefs}

\newcommand\cyr
{
\renewcommand\rmdefault{wncyr}
\renewcommand\sfdefault{wncyss}
\renewcommand\encodingdefault{OT2}
\normalfont
\selectfont
}
\DeclareTextFontCommand{\textcyr}{\cyr}

\DeclareMathAlphabet{\mathpzc}{OT1}{pzc}{m}{it}

\newcommand{\la}{\label} 

\newcommand{\al}{\alpha}

\def \d{\rm d}
\newcommand{\p}{\partial}
\newcommand{\bs}{\boldsymbol}
\newcommand{\rar}{\rightarrow}

\begin{document}

\title{Piecewise linear constitutive relations for stretch-limited elastic strings}
\author{R. Bustamante}
\author{K. R. Rajagopal}
\author{C. Rodriguez}
\thanks{K. R. Rajagopal and C. Rodriguez gratefully acknowledge support from NSF DMS-2307562.}

\begin{abstract}
This study proposes a simple and novel class of stretch-limiting constitutive relations for perfectly flexible elastic strings drawn from modern advances in constitutive theory for elastic bodies. We investigate strings governed by constitutive relations where stretch is a bounded, piecewise linear function of tension, extending beyond the traditional Cauchy elasticity framework. Our analysis includes explicit solutions for both catenaries and longitudinal, piecewise constant stretched motions.
\end{abstract}

\maketitle

\section{Introduction}

In the past 20 years, the term ``elastic" has experienced a significant evolution, expanding its definition and scope. Unlike Cauchy elasticity \cite{Cauchy1822, Cauchy1828}, wherein the stress is given explicitly as a function of the deformation gradient, or Green elasticity  \cite{Green1839} where one assumes the existence of a stored energy in terms of the deformation gradient, the second listed author \cite{Raj03} suggested an implicit relationship between the density, the stress, and the deformation gradient. The second listed author and co-workers have shown that the ambit of elastic bodies, namely bodies incapable of dissipation (converting mechanical working into heat), is far larger than that of Cauchy elastic bodies and includes bodies described by such implicit constitutive relations which do not belong to the class of Cauchy elastic bodies (see \cite{Raj03, Raj07, Raj11, RajSri07, RajSri09, BustRaj20, BustRaj21, RajSacc22}). A special sub-class of such implicit elastic bodies are isotropic elastic bodies wherein the right Cauchy-Green tensor is a function of the stress and density (or isotropic bodies wherein the left Cauchy-Green tensor is a function of the stress and density). When displacement gradients are restricted to be small, we obtain the linearized strain as a function of the stress and the density. In general, this relationship between the linearized strain and the stress can be nonlinear, a sharp contrast to classical constitutive theory that only produces linear relations between linearized strain and stress. Recent studies expanding these theories into viscoelastic and dispersive settings can be found in, e.g., \cite{ErbaySengul15, ErbaySengul20, Erbayetal20, Erbayetal24, Sengul21, Sengul21b, Sengul22}.

This paper focuses on the physically important but simpler setting in which classical and modern advances in constitutive theory for elastic bodies can be explored: \textit{perfectly flexible strings}. Interest in the study of the mechanics of strings and fibers has a storied history; it attracted the attention of da Vinci, Galileo, Euler, the Bernoullis, Savart, Lagrange, Leibniz, Ricatti, Huygens, Navier, Poisson, Lam\'e and other distinguished scientists. The interested reader can find a detailed account of the history of investigations into the mechanics of strings/fibers/wires in the work of Todhunter \cite{todhunter1886}. All of these early studies were within the context of special theories and not within the purview of general nonlinear theories of elasticity. Studies within the context of nonlinear theories are too numerous to mention and one can find copious references to both recent and older studies mentioned above in the treatise by Antman \cite{antman2004}.

For a perfectly flexible string, a constitutive relation is specified using only two scalar quantities. The frame indifferent kinematic variable of a string is the stretch $\nu = |\p_s \bs r|$ of the curve in space,
$$s \ni [0,L] \mapsto \bs r(s,t),$$
defining its configuration at time $t$. The contact force $\bs n$ is assumed to be proportional to the unit tangent to the curve defining the configuration of the string, $\bs n = N \nu^{-1}\p_s \bs r$. Thus, there is only one stress variable, the tension $N$. Classical constitutive relations take the form 
\begin{align}
N = \hat N(\nu), \label{eq:classicalelastic}
\end{align} 
in keeping with the traditional approach of Cauchy elasticity wherein one specifies the stress variable in terms of the deformation gradient, a kinematic variable (see Section 2 for more details).

Recent works have re-examined the mechanics of perfectly flexible strings within the modern framework introduced in \cite{Raj03}. The works \cite{Rod21, Rod22} initiated the study of elastic \textit{stretch-limited} strings satisfying constitutive relations of the form 
\begin{align}
	\nu = \hat \nu(N), 
\end{align}
where $\hat \nu : \mathbb R \rightarrow [\nu_0, \nu_1]$ is a bounded, non-decreasing, function, and there exist $N_0, N_1 > 0$ such that 
\begin{align}
\hat \nu(N) = 
\begin{cases}
	\nu_0 &\mbox{ if } N < -N_0, \\
	\nu_1 &\mbox{ if } N > N_1.
\end{cases}
\end{align} 
In \cite{Rod21}, the author established results on the existence and { multiplicity of equilibria} for stretch-limited catenaries, strings suspended between two points, a problem that attracted the interests of the likes of Galileo, Bernoulli, Huygens, Leibniz, among others. He found that the number of solutions for such stretch-limited strings deviates significantly from the case of traditional classical elastic strings satisfying \eqref{eq:classicalelastic}. The work \cite{Rod22} studied the propagation of longitudinal waves in a semi-infinite stretch-limited string with the finite end held fixed and the other end asymptotically in tension. There, he proved the orbital asymptotic stability of an explicit two-parameter family of piecewise constant stretched motions.

The latter two authors recently explored Cosserat rods within frameworks arising from modern advances in constitutive theory, considering both elasticity and viscoelasticity. In \cite{RajRod23a}, they studied Cosserat rods satisfying strain-limiting constitutive relations between the rod's strains and stress relevant to modeling biological fibers such as DNA. They derived explicit equilibrium states for such rods that exhibited a variety of rich characteristics including shearing instabilities and Poynting effects. In \cite{RajRod22} they studied the response of an inextensible, unshearable, viscoelastic rod that moves on a plane and possesses an evolving natural configuration characterized by its curvature (see \cite{Raj95} for a discussion of the notion of natural configuration, a concept studied in detail earlier by Eckart \cite{Eckart48}; see also \cite{RajSri00} for the connection between the evolution of the natural configuration and the rate of entropy production). They established global well-posedness and asymptotic dynamics for solutions of the governing quasi-static equations of motion. The model studied in \cite{RajRod22} was extended in \cite{RajRod23b} to encompass Cosserat rods moving in three-dimensional space that possess rate-dependent evolving natural configurations. In an interesting departure from the usual imposition of the second law of thermodynamics in a point-wise strong sense, the work \cite{RajRod23b} enforces a weaker requirement, namely that the Clausius-Duhem inequality hold over the whole body in an integral sense. They showed that one can have two possible constitutive relations, one that meets the stricter point-wise inequality and another that meets the global integral inequality.

In this work, we introduce a simple class of stretch-limiting, piecewise linear constitutive relations for elastic strings of the form $\nu = \hat \nu(N)$. The organization of the paper is as follows. In Section 2, we briefly review the kinematics, balance laws and constitutive theory governing perfectly flexible elastic strings. In Section 3, we present a simple form of $\hat \nu(N)-1$ that limits the amount of compression and extension the string can undergo, is linear in $N$ when the string is extensible, and has different tangent moduli (i.e., proportionality constants) for extension versus compression (see \eqref{eq20}). In Section 4, we formulate simple boundary value problems, which are then solved from a semi-inverse perspective in Section 5 (for catenaries) and Section 6 (for piecewise constant stretched motions). Finally, in Section 7, we discuss potential applications of the proposed model. 

{ The problems solved in Sections 5 and 6 illustrate key differences between the present work and previous studies while also offering potential applications. Section 5 provides explicit solutions for catenaries, unlike the more abstract results on the existence and multiplicity of solutions obtained in \cite{Rod21}. Similarly, the explicit piecewise constant stretch motions obtained in Section 6 involve a compressed segment and an elongated segment with distinct tangent moduli, whereas the solutions in \cite{Rod22} involve a maximally stretched segment and an elongated but still flexible segment. One application of the solutions in Section 5 is in modeling quasi-static deformations of stretch-limited strings, which may be relevant in civil engineering. In such cases, allowing for some elasticity while imposing a stretch limit offers insight into how these strings behave differently from the standard inextensible assumption often used in structural analysis (see, for example, Section 5/8 of \cite{MeriamKraige}). The solutions in Sections 5 and 6 also serve as useful benchmarks for numerical methods, which are essential for studying more realistic problems involving these new constitutive theories. Future work will explore some of these numerical methods. In summary, the solutions obtained in this work offer new concrete perspectives on stretch-limited strings and open the door to further analytical and computational developments.}

\section{Perfectly flexible strings}
In this section, we provide a brief summary of the kinematics and balance laws relevant to the modeling of perfectly flexible strings. A careful and detailed description of the same can be found in the treatise \cite{antman2004}. 

\subsection{Kinematics}
Let $\mathbb E^3$ be three-dimensional Euclidean space. We identify its associated translation space with $\mathbb R^3$ via a fixed, right-handed, orthonormal basis $\{ \bs i, \bs j, \bs k \}$.

Let $L > 0$ be the reference length of the string which we assume coincides with the undeformed length of the string. We parameterize the \emph{material points} of the string by the interval $[0,L]$. The \emph{current configuration} of the string at time $t$ is parameterized by a curve $$[0,L] \ni s \mapsto \bs r(s,t) \in \mathbb E^3.$$ The tangent to the curve $\bs r(\cdot, t)$ at $s$ is $\bs r_s(s,t) := \p_s \bs r(s,t)$, and the velocity of the material point $s$ is $\bs r_t(s,t) := \p_t \bs r(s,t)$. The \emph{stretch} $\nu(s,t)$ of the string at $(s,t)$ is 
\begin{align*}
	\nu(s,t) := |\bs r_s(s,t)|.
\end{align*}
We require that the configuration is regular so that $\nu > 0$ throughout the motion of the string. 

\subsection{Balance laws and constitutive theory}

The \emph{classical equations of motion} expressing local balance of linear momentum for a string are given by: 
\begin{align}
	(\rho A)(s) \bs r_{tt}(s,t) = \bs n_s(s,t) + \bs f(s,t), \quad (s,t) \in I \times [0,T]. \label{eq:motion}
\end{align}
Here $(\rho A)(s)$ is the mass per unit reference length at $s$, $\bs f(s,t)$ is an external body force per unit reference length at $(s,t)$, and $\bs n(s,t)$ is the contact force at $(s,t)$. 

The defining characteristic of a perfectly flexible string is that the contact force is proportional to the unit tangent of the string's configuration: there exists a scalar-valued function $N(s,t)$, the \emph{tension}, such that 
\begin{align}
	\bs n(s,t) = N(s,t) \frac{\bs r_s(s,t)}{|\bs r_s(s,t)|}. \label{eq:stringcontact}
\end{align} 
We note that $N$ satisfies 
\begin{align}
		\la{eqN19}N(s,t) =\mbox{sgn}(\boldsymbol{r}_s(s,t)\cdot\boldsymbol{n}(s,t))|\boldsymbol{n}(s,t)|.
\end{align}

The mechanical properties of the string are modeled by specifying a relation between the stretch $\nu$ and tension $N$. Classically, a string is said to be \emph{elastic} if there exists a function $\hat N(\nu,s)$ such that 
\begin{align}
	N(s,t) = \hat N(\nu(s,t), s), \label{eq:constN}
\end{align}
with $\hat N(1,s) = 0$, $\nu \mapsto \hat N(\nu,s)$ increasing, $\lim_{\nu \rar 0} \hat N(\nu,s) = -\infty$ and $\lim_{\nu \rightarrow \infty} \hat N(\nu,s) = \infty$ for all $s$. These mathematical restrictions on $\hat N$ reflect the physical assumptions:
\begin{itemize} 
\item A configuration of the string is natural ($N = 0$) if and only if it is unstretched ($\nu = 1$).
\item An increase in tension coincides with an increase in stretch.
\item Total compression ($\nu = 0$) requires infinite compressive force and infinite stretch requires infinite tensile force.  
\end{itemize}
The previous mathematical assumptions imply that $\hat N(\cdot,s)$ has an inverse function $\hat \nu(\cdot, s)$, and the constitutive relation between stretch and tension can be recast as 
\begin{align}
	\nu(s,t) = \hat \nu(N(s,t),s). \label{eq:const}
\end{align}

\section{A new constitutive relation for elastic strings \la{implicit}}

In this section we propose a new constitutive relation for perfectly flexible elastic strings of the form:
\begin{align}
	\nu = \hat \nu(N),
\end{align}
where we have suppressed the dependence of the variables on $s$ and $t$.
As discussed in the previous section, $\nu > 0$ is the stretch and $N\in\mathbb{R}$ is the tension of the string. We will impose the following weaker restrictions than in the classical setting:
\begin{itemize}
\item A configuration of the string is natural ($N = 0$) if and only if it is unstretched ($\nu = 1$).
\item Tensile forces ($N > 0$) correspond to extended configurations ($\nu > 1$).
\item Compressive forces ($N < 0$) correspond to compressed configurations ($0 < \nu < 1$).
\end{itemize}

\subsection{A piecewise linear constitutive relation}

From the physical point of view, it is very reasonable that there are limits to the amount of compression and extension that a real string can endure before failure. In Figure \ref{fig1} on the left, we have a schematic depiction for $\nu-1$ versus $N$, while on the right we have a simplified model for $\hat{\nu}-1$ for the approximation of the above behavior.

\begin{figure}[!ht]
	\begin{center}
		\includegraphics[width=15cm]{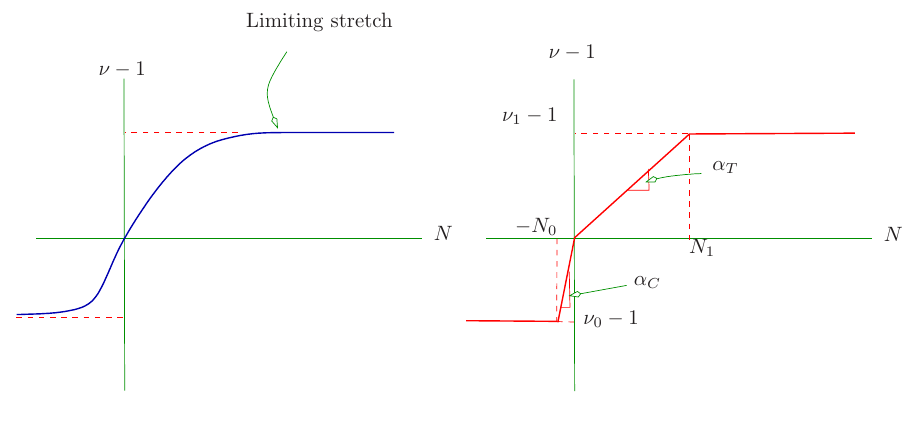}
	\end{center}
	\caption{Stretch limiting model for the string.\la{fig1}}
\end{figure}

The simplified constitutive relation that we propose in this work is the piecewise differentiable model:
\begin{equation}
\la{eq20}\hat{\nu}(N)-1=\left\{
\begin{array}{lll}
\nu_1-1\quad &\mbox{if }& N>N_1,\\
\alpha_TN &\mbox{if }& 0\le N\le N_1,\\
\alpha_CN &\mbox{if }& -N_0\leq N<0,\\
\nu_0-1\quad &\mbox{if }& N<-N_0.
\end{array}
\right.
\end{equation}
where $N_0$, $N_1$, and $0 < \nu_0 < 1 < \nu_1$ are fixed positive constants. If $N>N_1$, then the extension of the string has reached its limit $\nu_1$, while if $N<-N_0$, compression has reached its limit $\nu_0$. When $0\leq N\leq N_1$ (respectively, $-N_0\leq N<0$), extension (respectively, compression) and tension are proportional. We assume that $\alpha_T < \alpha_C$ so that the string can deform more with relatively small internal load when in compression. In particular, strings modeled by \eqref{eq20} have \emph{different tangent moduli} for extension versus compression.

It is important to note that the tension is not a function of the stretch, however, the stretch is a function of the tension. Thus, the constitutive relation under consideration does not belong to the class of classical Cauchy elastic bodies. 

\section{Boundary value problems \la{boundary}}

In this section, we set-up a boundary value problem for a uniform string satisfying the constitutive relation \eqref{eq20}. For certain calculations, we will use the alternative expression for the constitutive relation \eqref{eq:stringcontact}:
\begin{equation}
	\la{eqOPI}\boldsymbol{r}_s=\frac{\hat{\nu}(N)}{N}\boldsymbol{n}.
\end{equation}

We seek to solve the governing field equations
\begin{equation}
\la{eq22}(\rho A)\boldsymbol{r}_{tt}(s,t)=\boldsymbol{n}_s(s,t)+\boldsymbol{f}(s,t),\quad
|\boldsymbol{r}_s|=\nu=\hat{\nu}(N),\quad
\boldsymbol{n}=N\frac{\boldsymbol{r}_s}{|\boldsymbol{r}_s|},
\end{equation}
for $\bs r(s,t)$ and $\bs n(s,t)$ 
where $\boldsymbol{f}$ is a given external body force and $(\rho A)$ is a constant. For the above equations, we impose boundary and initial conditions
\begin{equation}
\la{eq23}\boldsymbol{r}(0,t)=\boldsymbol{0},\quad\boldsymbol{r}(L,t)=L\boldsymbol{k},\quad
\boldsymbol{r}(s,0)=\boldsymbol{u}(s),\quad\boldsymbol{r}_t(s,0)=\boldsymbol{v}(s),
\end{equation}
where
\begin{equation}
\boldsymbol{u}(0)=\boldsymbol{0},\quad
\boldsymbol{u}(L)=L\boldsymbol{k},\quad
\boldsymbol{v}(0)=\boldsymbol{0},\quad
\boldsymbol{v}(L)=\boldsymbol{0}. \label{eq24}
\end{equation}

\subsection{Alternative forms of the field equations}

In this section, we assume that $N \neq 0$. If $N > 0$, then $N = |\bs n|$, and \eqref{eqOPI} and \eqref{eqN19} imply that
\begin{equation}
\boldsymbol{r}_s=\hat{\nu}(|\boldsymbol{n}|)\frac{\boldsymbol{n}}{|\boldsymbol{n}|}.
\end{equation}
Thus, by \eqref{eq23},
\begin{equation}
\la{eq26}\boldsymbol{r}(s,t)=\int_0^s\hat{\nu}(|\boldsymbol{n}(\xi,t)|)\frac{\boldsymbol{n}(\xi,t)}{|\boldsymbol{n}(\xi,t)|}\thinspace{\d}\xi.
\end{equation}
Inserting \eqref{eq26} into \eqref{eq22}, we obtain
\begin{equation}
\la{eq27}(\rho A)\frac{\partial^2}{\partial t^2}\left\{\int_0^s\hat{\nu}(|\boldsymbol{n}(\xi,t)|)\frac{\boldsymbol{n}(\xi,t)}{|\boldsymbol{n}(\xi,t)|}\thinspace{\d}\xi\right\}=\frac{\partial\boldsymbol{n}}{\partial s}(s,t)+\boldsymbol{f}(s,t).
\end{equation}

In the case that $N < 0$, the only modification we need to consider to the above equation is to use $\hat{\nu}(-|\boldsymbol{n}(\xi,t)|)$ instead of $\hat{\nu}(|\boldsymbol{n}(\xi,t)|)$.

By \eqref{eq26}, the boundary condition \eqref{eq23}$_1$ is satisfied automatically, and \eqref{eq23}$_2$ becomes
\begin{equation}
\int_0^L\hat{\nu}(|\boldsymbol{n}(\xi,t)|)\frac{\boldsymbol{n}(\xi,t)}{|\boldsymbol{n}(\xi,t)|}\thinspace{\d}\xi=L\boldsymbol{k}.
\end{equation}
The initial conditions \eqref{eq23}$_{3,4}$ become
\begin{equation}
\int_0^s\hat{\nu}(|\boldsymbol{n}(\xi,0)|)\frac{\boldsymbol{n}(\xi,0)}{|\boldsymbol{n}(\xi,0)|}\thinspace{\d}\xi=\boldsymbol{u}(s),\quad
\frac{\partial}{\partial t}\left[\int_0^s\hat{\nu}(|\boldsymbol{n}(\xi,t)|)\frac{\boldsymbol{n}(\xi,t)}{|\boldsymbol{n}(\xi,t)|}\thinspace{\d}\xi\right]_{t=0}=\boldsymbol{v}(s).
\end{equation}

\subsection{Dimensionless equations}

We define the following dimensionless variables
\begin{equation}
\la{eqN30}\bar{s}=\frac{s}{L},\quad
\bar{t}=\sqrt{\frac{g}{L}}t,\quad
\bar{\boldsymbol{n}}=\alpha_T\boldsymbol{n},\quad
\bar{\boldsymbol{r}}=\frac{1}{L}\boldsymbol{r},\quad
\zeta=(\rho A)gL\alpha_T,\quad
\bar{\boldsymbol{f}}=L\alpha_T\boldsymbol{f},\quad
\bar{N}=\alpha_T N, 
\end{equation}
where $g$ is the gravitational constant. Then \eqref{eq22} and \eqref{eqN19} become
\begin{gather}
\zeta\frac{\partial^2\bar{\boldsymbol{r}}}{\partial\bar{t}^2}=\frac{\partial\bar{\boldsymbol{n}}}{\partial\bar{s}}+\bar{\boldsymbol{f}}, \label{eq31}\\
|\bar{\boldsymbol{r}}_{\bar{s}}|=\hat{\nu}(\bar{N}),\quad
\bar{\boldsymbol{n}}=\bar{N}\frac{\bar{\boldsymbol{r}}_{\bar{s}}}{|\bar{\boldsymbol{r}}_{\bar{s}}|},\quad
\bar{N}:=\mbox{sgn}(\bar{\boldsymbol{r}}_{\bar{s}}\cdot\bar{\boldsymbol{n}})|\bar{\boldsymbol{n}}|, \label{eq32}
\end{gather}
where $\bar{\boldsymbol{r}}_{\bar{s}}=\frac{\partial\bar{\boldsymbol{r}}}{\partial\bar{s}}$. The boundary and initial conditions \eqref{eq23}, \eqref{eq24} are
\begin{equation}
\bar{\boldsymbol{r}}(0,\bar{t})=\boldsymbol{0},\quad
\bar{\boldsymbol{r}}(1,\bar{t})=\boldsymbol{k},\quad
\bar{\boldsymbol{r}}(\bar{s},0)=\bar{\boldsymbol{u}},\quad
\bar{\boldsymbol{r}}_{\bar{t}}(\bar{s},0)=\bar{\boldsymbol{v}}, \label{eq33}
\end{equation}
where
\begin{equation}
\bar{\boldsymbol{u}}=\frac{1}{L}\boldsymbol{u},\quad
\bar{\boldsymbol{v}}=\frac{1}{\sqrt{Lg}}\boldsymbol{v},
\end{equation}
and $\bar{\boldsymbol{r}}_{\bar{t}}=\frac{\partial\bar{\boldsymbol{r}}}{\partial\bar{t}}$.

In the case of the alternative representation \eqref{eq26} we obtain (in the case $N>0$)
\begin{equation}
\bar{\boldsymbol{r}}(\bar{s},\bar{t})=
\int_0^{\bar{s}}\hat{\nu}(|\bar{\boldsymbol{n}}(\bar{\xi},\bar{t})|)\frac{\bar{\boldsymbol{n}}(\bar{\xi},\bar{t})}{|\bar{\boldsymbol{n}}(\bar{\xi},\bar{t})|}\thinspace{\d}\bar{\xi}, \label{eq32dimensionless}
\end{equation}
and from \eqref{eq27} we have
\begin{equation}
\la{eq36}\zeta\frac{\partial^2}{\partial\bar{t}^2}\left[\int_0^{\bar{s}}\hat{\nu}(|\bar{\boldsymbol{n}}(\bar{\xi},\bar{t})|)\frac{\bar{\boldsymbol{n}}(\bar{\xi},\bar{t})}{|\bar{\boldsymbol{n}}(\bar{\xi},\bar{t})|}\thinspace{\d}\bar{\xi}\right]
=\frac{\partial\bar{\boldsymbol{n}}}{\partial\bar{s}}+\bar{\boldsymbol{f}}.
\end{equation}
The boundary condition \eqref{eq33}$_2$ becomes
\begin{equation}
\la{eq37}\int_0^{1}\hat{\nu}(|\bar{\boldsymbol{n}}(\bar{\xi},\bar{t})|)\frac{\bar{\boldsymbol{n}}(\bar{\xi},\bar{t})}{|\bar{\boldsymbol{n}}(\bar{\xi},\bar{t})|}\thinspace{\d}\bar{\xi}=\boldsymbol{k},
\end{equation}
while for the initial conditions \eqref{eq33}$_{3,4}$, we have
\begin{eqnarray}
\la{eq38}\int_0^{\bar{s}}\hat{\nu}(|\bar{\boldsymbol{n}}(\bar{\xi},0)|)\frac{\bar{\boldsymbol{n}}(\bar{\xi},0)}{|\bar{\boldsymbol{n}}(\bar{\xi},0)|}\thinspace{\d}\bar{\xi}=\bar{\boldsymbol{u}}(\bar{s}),\\
\la{eq39}\frac{\partial}{\partial\bar{t}}\left[\int_0^{\bar{s}}\hat{\nu}(|\bar{\boldsymbol{n}}(\bar{\xi},\bar{t})|)\frac{\bar{\boldsymbol{n}}(\bar{\xi},\bar{t})}{|\bar{\boldsymbol{n}}(\bar{\xi},\bar{t})|}\thinspace{\d}\bar{\xi}\right]_{\bar{t}=0}=\bar{\boldsymbol{v}}(\bar{s}).
\end{eqnarray}

Finally, our model \eqref{eq20} becomes
\begin{equation}
\la{eqN40}\hat{\nu}(\bar{N})-1=\left\{
\begin{array}{lll}
\nu_1-1\quad &\mbox{if } \bar{N}>\bar{N}_1,\\
\bar{N} &\mbox{if } 0\le \bar{N}\le \bar{N}_1,\\
\frac{\alpha_C}{\alpha_T}\bar{N} &\mbox{if } -\bar{N}_0\le\bar{N}<0,\\
\nu_0-1\quad &\mbox{if } \bar{N}<-\bar{N}_0,
\end{array}
\right.
\end{equation}
where $\bar{N}_1=\alpha_TN_1$ and $\bar{N}_0=\alpha_TN_0$. 

Going forward we work with the dimensionless variables and equations derived in this section. \textbf{To decrease the amount of necessary notation, we will drop the over-bars. }

\section{Catenaries}

In this section, we assume that the different variables do not depend on time and obtain solutions to the boundary value problem from the semi-inverse perspective.

Since the variables are time independent, $\boldsymbol{r}_{tt}=\boldsymbol{0}$ and $\boldsymbol{r}_t=0$. From \eqref{eq31}, we obtain ${\boldsymbol{n}}_{{s}}=-{\boldsymbol{f}}$, and as a result,\footnote{We use the notation ${\boldsymbol{n}}_{{s}}=\frac{\partial{\boldsymbol{n}}}{\partial{s}}$ and  we recall ${\boldsymbol{r}}_{{s}}=\frac{\partial{\boldsymbol{r}}}{\partial{s}}$. }
\begin{equation}
\la{eq40}{\boldsymbol{n}}=-\int_0^{{s}}{\boldsymbol{f}}({\xi})\thinspace{\d}{\xi}+{\boldsymbol{n}}_0,
\end{equation}
where ${\boldsymbol{n}}_0$ is a constant dimensionless vector. Inserting \eqref{eq40} into \eqref{eq32} and assuming that for $0\le {s}\le 1$ we have ${\boldsymbol{r}}_{{s}}\cdot{\boldsymbol{n}}\ge 0$, we obtain
\begin{equation}
\la{eq41}{\boldsymbol{r}}_{{s}}(s)=\frac{\hat{\nu}\left(|{\boldsymbol{n}}_0-\int_0^{{s}}{\boldsymbol{f}}({\xi})\thinspace{\d}{\xi}|\right)}
{\left|{\boldsymbol{n}}_0-\int_0^{{s}}{\boldsymbol{f}}({\xi})\thinspace{\d}{\xi}\right|}
\left\{{\boldsymbol{n}}_0-\int_0^{{s}}{\boldsymbol{f}}({\xi})\thinspace{\d}{\xi}\right\}.
\end{equation}
As a result, we obtain
\begin{equation}
\la{eq42}{\boldsymbol{r}}(s)=\int_0^{{s}}
\frac{\hat{\nu}\left(|{\boldsymbol{n}}_0-\int_0^{{\xi}}{\boldsymbol{f}}({\eta})\thinspace{\d}{\eta}|\right)}
{\left|{\boldsymbol{n}}_0-\int_0^{{\xi}}{\boldsymbol{f}}({\eta})\thinspace{\d}{\eta}\right|}
\left\{{\boldsymbol{n}}_0-\int_0^{{\xi}}{\boldsymbol{f}}({\eta})\thinspace{\d}{\eta}\right\}\thinspace{\d}{\xi}
+{\boldsymbol{r}}_0.
\end{equation}
From the boundary conditions \eqref{eq33} we have ${\boldsymbol{r}}(0)=\boldsymbol{0}$ that implies that ${\boldsymbol{r}}_0=\boldsymbol{0}$, and from ${\boldsymbol{r}}(1)=\boldsymbol{k}$ we can obtain ${\boldsymbol{n}}_0$.

We consider the particular case of the force due to gravity (\textit{catenaries}) which is equivalent to ${\boldsymbol{f}}=-\zeta\boldsymbol{e}_v$ where $\boldsymbol{e}_v$ is the unit vector in the vertical direction. The relation \eqref{eq42} and the fact that $\bs r_0 = \bs 0$ imply
\begin{equation}
\la{eq43}{\boldsymbol{r}}({s})=\int_0^{{s}}\frac{\hat{\nu}\left(|{\boldsymbol{n}}_0+{\xi}\zeta\boldsymbol{e}_v|\right)}
{\left|{\boldsymbol{n}}_0+{\xi}\zeta\boldsymbol{e}_v\right|}\left\{{\boldsymbol{n}}_0+{\xi}\zeta\boldsymbol{e}_v\right\}\thinspace{\d}{\xi}.
\end{equation}

\subsection{The case of tensile internal forces and inextensibility is not reached.}

To advance in the analysis of the boundary value problem, let us assume that $\boldsymbol{k}=\boldsymbol{e}_1$ and that $\boldsymbol{e}_v=\boldsymbol{e}_2$, as depicted in Figure \ref{fig2}.

\begin{figure}[!ht]
	\begin{center}
		\includegraphics[width=10cm]{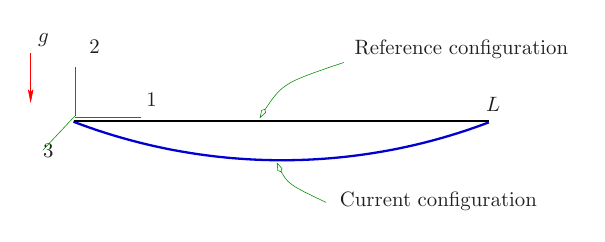}
	\end{center}
	\caption{A string under the force of gravity. \la{fig2}}
\end{figure}

We use the constitutive relation \eqref{eqN40}, and in this first case, we assume that $N=({\boldsymbol{r}}_{{s}}\cdot{\boldsymbol{n}})|{\boldsymbol{n}}|$ is such that $0\le {N}\le {N}_1$ for $0\le {s}\le 1$. Then from \eqref{eqN40} and \eqref{eq43} we obtain
\begin{equation}
{\boldsymbol{r}}(s)=\int_0^{{s}}\left(1+\frac{1}{\left|{\boldsymbol{n}}_0+{\xi}\zeta\boldsymbol{e}_2\right|}\right)\left\{{\boldsymbol{n}}_0+{\xi}\zeta\boldsymbol{e}_2\right\}\thinspace{\d}{\xi}.
\end{equation}
For each component ${r}_i$ of ${\boldsymbol{r}}$ (the components of ${\boldsymbol{n}}_0$ are ${n}_{0,i}$), we obtain
\begin{eqnarray}
{r}_1({s})&=&{n}_{0,1}\ell({s}),\quad {r}_3({s})={n}_{0,3}\ell({s}),\\
{r}_2({s})&=&\frac{1}{2\zeta}\left[
2\sqrt{{n}_{0,1}^2+\left({n}_{0,2}+\zeta{s}\right)^2+{n}_{0,3}^2}
+\zeta{s}\left(2{n}_{0,2}+\zeta{s}\right)
-2\sqrt{{n}_{0,1}^2+{n}_{0,2}^2+{n}_{0,3}^2}\right],
\end{eqnarray}
where we have defined
\begin{equation}
\ell({s})={s}+\frac{1}{\zeta}\ln\left[
\frac{{n}_{0,2}+\zeta{s}+\sqrt{{n}_{0,1}^2+\left({n}_{0,2}+\zeta{s}\right)^2+{n}_{0,3}^2}}
{{n}_{0,2}+\sqrt{{n}_{0,1}^2+{n}_{0,2}^2+{n}_{0,3}^2}}\right].
\end{equation}

From the boundary condition \eqref{eq33}$_2$, we have ${\boldsymbol{r}}(1)=\boldsymbol{e}_1$. Then
\begin{equation}
{r}_1(1)=1,\quad
{r}_2(1)=0,\quad
{r}_3(1)=0,
\end{equation}
which can be used to find ${n}_{0,1}$, ${n}_{0,2}$ and ${n}_{0,3}$.

\subsection{The case of tensile external forces and inextensibility is reached for a part of the string. \la{sec312}}

In this section we study the case when there exist $\tilde{{s}}$ and $\breve{{s}}$ such that $0<\tilde{{s}}<1-\breve{{s}}<1$ and
\begin{eqnarray}
{N}>{N}_1\quad&&\mbox{if}\quad 0\le{s}\le \tilde{{s}},\\
0\le{N}\le{N}_1\quad&&\mbox{if}\quad \tilde{{s}}<{s}<1-\breve{{s}},\\
{N}>{N}_1\quad&&\mbox{if}\quad 1-\breve{{s}}<{s}< 1,
\end{eqnarray}
i.e., two parts of the string reach inextensibility.

In the case $0\le{s}\le \tilde{{s}}$ since ${N}>{N}_1$ from \eqref{eqN40} we have $\hat{\nu}({N})=\nu_1$ and from \eqref{eq43} we have
\begin{equation}
{\boldsymbol{r}}({s})=\int_0^{{s}}\frac{\nu_1}
{\left|{\boldsymbol{n}}_0+{\xi}\zeta\boldsymbol{e}_2\right|}\left\{{\boldsymbol{n}}_0+{\xi}\zeta\boldsymbol{e}_2\right\}\thinspace{\d}{\xi},
\end{equation}
and for the components of ${\boldsymbol{r}}$ we obtain
\begin{eqnarray}
{r}_1({s})&=&{n}_{0,1}\ell_A({s}),\quad {r}_3({s})={n}_{0,3}\ell_A({s}),\\
{r}_2({s})&=&\frac{\nu_1}{\zeta}\left[\sqrt{{n}_{0,1}^2+\left({n}_{0,2}+\zeta{s}\right)^2+{n}_{0,3}^2}
-\sqrt{{n}_{0,1}^2+{n}_{0,2}^2+{n}_{0,3}^2}\right],
\end{eqnarray}
where we have defined
\begin{equation}
\ell_A({s})=\frac{\nu_1}{\zeta}\ln\left[
\frac{{n}_{0,2}+\zeta{s}+\sqrt{{n}_{0,1}^2+\left({n}_{0,2}+\zeta{s}\right)^2+{n}_{0,3}^2}}
{{n}_{0,2}+\sqrt{{n}_{0,1}^2+{n}_{0,2}^2+{n}_{0,3}^2}}
\right].
\end{equation}

In the case $\tilde{{s}}<{s}<1-\breve{{s}}$ from \eqref{eqN40} we have $\hat{\nu}({N})={N}+1$ and from \eqref{eq43} we get
\begin{equation}
{\boldsymbol{r}}=\int_{\tilde{{s}}}^{{s}}\left(1+\frac{1}{\left|{\boldsymbol{n}}_0+{\xi}\zeta\boldsymbol{e}_2\right|}\right)\left\{{\boldsymbol{n}}_0+{\xi}\zeta\boldsymbol{e}_2\right\}\thinspace{\d}{\xi}
+{\boldsymbol{r}}_0,
\end{equation}
and for each component of ${\boldsymbol{r}}$ we obtain
\begin{eqnarray}
{r}_1({s})&=&{n}_{0,1}\ell_B({s})+{r}_{0,1},\quad {r}_3({s})={n}_{0,3}\ell_B({s})+{r}_{0,3},\\
{r}_2({s})&=&\frac{1}{2\zeta}\left[2\sqrt{{n}_{0,1}^2+\left({n}_{0,2}+\zeta{s}\right)^2+{n}_{0,3}^2}
+\zeta{s}\left(2{n}_{0,2}+\zeta{s}\right)\right.\nonumber\\
&&\left.-2\sqrt{{n}_{0,1}^2+\left({n}_{0,2}+\zeta\tilde{{s}}\right)^2+{n}_{0,3}^2}
-\zeta\tilde{{s}}\left(2{n}_{0,2}+\zeta\tilde{{s}}\right)\right]+{r}_{0,2},
\end{eqnarray}
where
\begin{equation}
\ell_B({s})={s}-\tilde{{s}}+\frac{1}{\zeta}\ln\left[
\frac{{n}_{0,2}+\zeta{s}+\sqrt{{n}_{0,1}^2+\left({n}_{0,2}+\zeta{s}\right)^2+{n}_{0,3}^2}}
{{n}_{0,2}+\zeta\tilde{{s}}+\sqrt{{n}_{0,1}^2+\left({n}_{0,2}+\zeta\tilde{{s}}\right)^2+{n}_{0,3}^2}}\right].
\end{equation}

For the case $1-\breve{{s}}<{s}< 1$ we again have ${N}>{N}_1$ thus $\hat{\nu}({N})=\nu_1$, and as a result from \eqref{eq43} we have
\begin{equation}
{\boldsymbol{r}}({s})=\int_{1-\breve{{s}}}^{{s}}\frac{\nu_1}
{\left|{\boldsymbol{n}}_0+{\xi}\zeta\boldsymbol{e}_2\right|}\left\{{\boldsymbol{n}}_0+{\xi}\zeta\boldsymbol{e}_2\right\}\thinspace{\d}{\xi}+\hat{{\boldsymbol{r}}}_0,
\end{equation}
from which we get
\begin{eqnarray}
{r}_1({s})&=&{n}_{0,1}\ell_C({s})+\hat{{r}}_{0,1},\quad {r}_3({s})={n}_{0,3}\ell_C({s})+\hat{{r}}_{0,3},\\
{r}_2({s})&=&\frac{\nu_1}{\zeta}\left\{\sqrt{{n}_{0,1}^2+\left({n}_{0,2}+\zeta{s}\right)^2+{n}_{0,3}^2}
-\sqrt{{n}_{0,1}^2+\left[{n}_{0,2}+\zeta\left(1-\breve{{s}}\right)\right]^2+{n}_{0,3}^2}\right\}+\hat{{r}}_{0,2},
\end{eqnarray}
where
\begin{equation}
\ell_C({s})=\frac{\nu_1}{\zeta}\ln\left\{
\frac{{n}_{0,2}+\zeta{s}+\sqrt{{n}_{0,1}^2+\left({n}_{0,2}+\zeta{s}\right)^2+{n}_{0,3}^2}}
{{n}_{0,2}+\zeta{s}+\sqrt{{n}_{0,1}^2+\left[{n}_{0,2}+\zeta\left(1-\breve{{s}}\right)\right]^2+{n}_{0,3}^2}}
\right\}.
\end{equation}

In the above results for ${r}_i({s})$ we have the following list of 11 unknown parameters: $\tilde{{s}}$, $\breve{{s}}$, ${r}_{0,1}$, ${r}_{0,2}$, ${r}_{0,3}$, $\hat{{r}}_{0,1}$, $\hat{{r}}_{0,2}$, $\hat{{r}}_{0,3}$, ${n}_{0,1}$, ${n}_{0,2}$ and ${n}_{0,3}$. These can be found from the 6 continuity conditions
\begin{equation}
\la{eqKiUk}{r}_i(\tilde{{s}}^-)={r}_i(\tilde{{s}}^+),\quad
{r}_i(1-\breve{{s}}^-)={r}_i(1-\breve{{s}}^+),\quad i=1,2,3,
\end{equation}
where the superscripts $-$ and $+$ mean approaching from the left and the right, respectively. As well as this we have the boundary conditions \eqref{eq33}$_2$, which in our case means
\begin{equation}
{r}_1(1)=1,\quad
{r}_2(1)=0,\quad
{r}_3(1)=0. \label{eq:rboundary}
\end{equation}
Finally, from \eqref{eqN40} we have that if ${s}=\tilde{{s}}$ and ${s}=1-\breve{{s}}$ we have that ${N}={N}_1$, which is equivalent to
\begin{equation}
{n}_{0,1}^2+\left({n}_{0,2}+\zeta\tilde{{s}}\right)^2+{n}_{0,3}^2={N}_1^2, \quad 
{n}_{0,1}^2+\left[{n}_{0,2}+\zeta\left(1-\breve{{s}}\right)\right]^2+{n}_{0,3}^2={N}_1^2. \label{eqRRtT}
\end{equation}
From \eqref{eqKiUk}, \eqref{eq:rboundary}, and \eqref{eqRRtT}, we have 11 equations determining the list of 11 unknowns. 

 \section{Motions with piecewise constant stretch}

In this section we consider the motion of a straight string $\bs r(s,t) = \chi(s,t) \bs e_1$ satisfying \eqref{eqN40}. We assume that the string has piecewise constant stretch separated by a shock front $\sigma(t) \in [0,1]$: there exist $\nu_- > 0$ and $\nu_+ > 0$ with $\nu_- \not = \nu_+$ such that 
\begin{align}
	\chi_s(s,t) = \begin{cases}
		\nu_- &\quad \mbox{ if } s < \sigma(t), \\
		\nu_+ &\quad \mbox{ if } s > \sigma(t). 
	\end{cases}
\end{align}
In this setting, we do not prescribe the placement of $\bs r(1,t)$ but, instead, the terminal tension $N_+ := N(1,t)$ is given. 

We will assume that the terminal tension is a tensile force $N_+ > 0$ and the string is separated into two extensible parts: a compressed part parameterized by $[0,\sigma(t)]$ and an elongated part parameterized by $[\sigma(t),1]$. In particular, it follows that 
\begin{align}
	\nu_0 < \nu_- = \frac{\alpha_C}{\alpha_T}N_- + 1 < 1 < \nu_+ = N_+ + 1 < \nu_1,
\end{align}    
where we have used \eqref{eqN40} to relate stretches to the (constant) tensions to the left and right of $\sigma(t)$. The case that both parts are elongated with one part of the string reaching inextensibility was treated in \cite{Rod22}.  

\subsection{Solution of the shock problem}

We seek configurations with piecewise constant stretch of the form  
\begin{align}
	\chi(s,t) = 
	\begin{cases}
	(\frac{\alpha_C}{\alpha_T} N_- + 1) s &\mbox{ if } s \in [0,\sigma(t)], \\
	(N_++1)(s-s_0) + v_+ t + \sigma_{0} &\mbox{ if } s \in [\sigma(t), 1].
	\end{cases}\label{eq:piecwise}
\end{align}
Here, $\sigma_0 \in (0,1)$ is the initial position of the shock front with reference position $s_0 \in (0,1)$, and $v_+ \neq 0$ is the (dimensionless) velocity of the particles comprising the elongated segment. $N_+ \in (0,N_1)$ is the terminal tension and $N_- \in (-N_0, 0)$ is the unknown constant tension in the compressed part of the string. 

We observe that the boundary conditions, $\chi(0,t) = 0$ and $N(1,t) = \chi_s(1,t) - 1 = N_+ < N_1$, are automatically satisfied, and the equations of motion hold on the segments $[0,\sigma(t)]$ and $[\sigma(t),1]$. The Rankine-Hugoniot jump conditions guaranteeing that $\bs r(s,t) = \chi(s,t) \bs e_1$ is a weak solution to \eqref{eq31} reduce to 
\begin{align}
	\llbracket N \rrbracket + \zeta \dot \sigma \llbracket \p_t \chi \rrbracket = 0. \label{eq:RH}
\end{align}
See, e.g., \cite{antman2004}. Above, $\llbracket y \rrbracket(t) = y(\sigma^+(t),t) - y(\sigma^-(t),t)$ is the jump across $s = \sigma(t)$ at time $t$. 

We now solve \eqref{eq:RH}. Continuity of $\chi$ requires that $\chi(\sigma(t)^-,t) = \chi(\sigma(t)^+,t)$ and differentiating in time yields
\begin{align}
 \sigma' \chi_s(\sigma^-,t) + \chi_t(\sigma^-,t) = \sigma' \chi_s(\sigma^+,t) + \chi_t(\sigma^+,t)  \iff	\sigma' = v_+ \Bigl (\frac{\alpha_C}{\alpha_T}N_- - N_+ \Bigr )^{-1}. \label{eq:jump2}
\end{align}
Since $\llbracket \p_t \chi \rrbracket = v_+$, \eqref{eq:jump2} implies that \eqref{eq:RH} is equivalent to
\begin{align}
	(N_- - N_+)\Bigl (N_- - \frac{\alpha_T}{\alpha_C} N_+\Bigr ) = \zeta v_+^2 \frac{\alpha_T}{\alpha_C}, 
\end{align} 
The assumption that the segment parameterized by $[0,\sigma(t)]$ is extensible and compressive, $-N_0 < N_- < 0$, implies that 
\begin{gather}
	N_- = \frac{1}{2}\Bigl (1 + \frac{\al_T}{\al_C} \Bigr ) N_+ - \Bigl [
	\frac{1}{4} \Bigl ( 1- \frac{\alpha_T}{\alpha_C} \Bigr )^2 N_+^2 + \zeta v_+^2 \frac{\al_T}{\al_C}
	\Bigr ]^{1/2}, \label{eq:lefttension}
\end{gather}
with
\begin{align}
\frac{N_+^2}{\zeta} < v_+^2 < \frac{N_+^2}{\zeta} + \frac{1}{\zeta} \Bigl ( \frac{\al_C}{\al_T} + 1 \Bigr )N_+ N_0 + \frac{\al_C}{\zeta \al_T} N_0^2. \label{eq:vconditions}
\end{align}
Then the velocity of the shock front can be written as 
\begin{align}
	\sigma' = v_+ \Bigl \{ \frac{1}{2}\Bigl (\frac{\al_C}{\al_T} - 1 \Bigr )N_+ - 
	\Bigl [\frac{1}{4}\Bigl ( \frac{\al_C}{\al_T} - 1 \Bigr )^2 N_+^2 + \zeta \frac{\al_C}{\al_T} v_+^2 \Bigr ]^{1/2}
	\Bigr \}^{-1}. \label{eq:shockspeed}
\end{align}
In conclusion, the Rankine-Hugoniot jump condition \eqref{eq:RH} is satisfied by $\chi$ with $N_- \in (-N_0, 0)$ if and only if $N_-$ is given by \eqref{eq:lefttension}, the inequalities \eqref{eq:vconditions} hold, and the shock speed is given by \eqref{eq:shockspeed}. 

The dimensionless speed of propagation in the elongated section is $c_+ = 1$ while the dimensionless speed of propagation in the compressed part is $c_- = \frac{\alpha_T}{\alpha_C} < 1 = c_+$. Consequently, we must have $v_+ > 0$ and
\begin{align}
-1 < \sigma' < -\frac{\al_T}{\al_C} \label{eq:Lax}
\end{align}
for Lax's geometric shock inequalities to hold \cite{Lax1957}. These conditions ensure the structurally stability of \eqref{eq:piecwise} as a weak solution to \eqref{eq31} \cite{Lax1957, Majda1984}. By \eqref{eq:shockspeed} and \eqref{eq:vconditions}, the condition $v_+ > 0$ implies that elongated section parameterized by $[\sigma(t),1]$ is \textit{growing} in time.

We now determine the implications of \eqref{eq:Lax}. Using \eqref{eq:shockspeed}, we see that $\sigma' > -1$ if and only if 
\begin{gather}
	\Bigl [ v_+ + \frac{1}{2}\Bigl (\frac{\al_C}{\al_T}  - 1\Bigr )N_+\Bigr ]^2 < \frac{1}{4}\Bigl( \frac{\al_C}{\al_T} + 1 \Bigr )^2 N_+^2 + \zeta v_+^2 \frac{\al_C}{\al_T} \iff \\
	 \Bigl ( 1 - \zeta \frac{\al_C}{\al_T} \Bigr )v_+^2 + \Bigl ( \frac{\al_C}{\al_T} - 1 \Bigr ) N_+ v_+ < 0.
\end{gather}
Thus, in order for the above solutions to satisfy Lax's shock inequality $\sigma' > -1$, we must have that $1-\zeta \frac{\al_C}{\al_T} < 0$ and 
\begin{align}
\Bigl ( \zeta \frac{\al_C}{\al_T} - 1 \Bigr )^{-1}\Bigl ( \frac{\al_C}{\al_T} - 1 \Bigr ) N_+ < v_+
\end{align}
The second condition $\sigma' < -\frac{\al_T}{\al_C}$ can similarly be seen to be equivalent to 
\begin{align}
	0 < \Bigl ( 1 - \zeta \frac{\al_T}{\al_C} \Bigr )v_+^2 + \Bigl (1 - \frac{\al_T}{\al_C} \Bigr ) N_+ v_+. \label{eq:lax2}
\end{align} 
If $1 - \zeta \frac{\al_T}{\al_C} \geq 0$, then \eqref{eq:lax2} holds automatically, but if $1 - \zeta \frac{\al_T}{\al_C} < 0$, then \eqref{eq:lax2} holds if and only if 
\begin{align}
	v_+ < \Bigl ( \zeta \frac{\al_T}{\al_C} - 1 \Bigr )^{-1} \Bigl (1 - \frac{\al_T}{\al_C} \Bigr ) N_+
\end{align}

In summary, piecewise constant stretch solutions of the form \eqref{eq:piecwise}, which satisfy \eqref{eq:lefttension}, \eqref{eq:vconditions}, and \eqref{eq:shockspeed}, verify Lax's geometric shock inequalities if and only if $v_+ > 0$, i.e., the elongated segment is growing in time. Additionally, the following conditions must be met: either
\begin{align}
 \frac{\al_T}{\al_C} < \zeta \leq \frac{\al_C}{\al_T} \quad \mbox{and} \quad \Bigl ( \zeta \frac{\al_C}{\al_T} - 1 \Bigr )^{-1}\Bigl ( \frac{\al_C}{\al_T} - 1 \Bigr ) N_+ < v_+, \label{eq:shock1}
\end{align}
or 
\begin{align}
	\zeta > \frac{\al_C}{\al_T} \quad \mbox{and} \quad \Bigl ( \zeta \frac{\al_C}{\al_T} - 1 \Bigr )^{-1}\Bigl ( \frac{\al_C}{\al_T} - 1 \Bigr ) N_+ < v_+ < \Bigl ( \zeta \frac{\al_T}{\al_C} - 1 \Bigr )^{-1} \Bigl (1 - \frac{\al_T}{\al_C} \Bigr ) N_+. \label{eq:shock2}
\end{align} 
In particular, if $\zeta = \frac{\al_T}{\al_C}[1 + \beta(\frac{\al_C^2}{\al_T^2}-1)]$, $\beta \in (0,1)$, $\frac{\al_C}{\al_T}$ and $\beta$ sufficiently close to 1, then 
\begin{align}
	\Bigl ( \zeta \frac{\al_C}{\al_T} - 1 \Bigr )^{-1}\Bigl ( \frac{\al_C}{\al_T} - 1 \Bigr ) N_+ < \frac{N_+}{\zeta^{1/2}}.
\end{align}
Therefore, \eqref{eq:vconditions} and \eqref{eq:shock1} are then satisfied for all $v_+$ sufficiently close to $\frac{N_+}{\zeta^{1/2}}$, establishing that the conditions \eqref{eq:vconditions} and either \eqref{eq:shock1} or \eqref{eq:shock2} are satisfied by an open set of values for the parameters $\frac{\al_C}{\al_T}$, $\zeta$, and $v_+$. 

\section{Final remarks \la{remarks}}

In the present work, we proposed a new type of piecewise linear constitutive relation for strings that limits the amount of extension and compression the string can undergo. Moreover, the tangent moduli relating tension to stretch are different for extension versus compression. We solved static and dynamic problems from the semi-inverse perspective with solutions to the latter having piecewise constant stretch. Numerical methods for solving more realistic static problems will be explored in future work, while dynamic problems will be studied using the method developed in \cite{bustamante2024}.

{The constitutive model developed in this paper provides a simple framework for describing the behavior of stretch-limited strings, which appear in various physical and biological settings and cannot be modeled using the classical approach. As shown in Figure 1 of \cite{Raj07}, in general, implicit constitutive theories are particularly useful in situations where the conventional approach of expressing the stress (or, in this case, the tension) explicitly as a function of the strain is inadequate. This is precisely the case for the constitutive relation \eqref{eq20} introduced here: when the stretch reaches a critical threshold, it remains fixed, yet the tension can continue to either decrease or increase without bound. This behavior fundamentally defies modeling within the classical framework, which expresses tension as a function of stretch. The model developed in this study has potential applications in biological systems, where stretch-limited strings arise in different contexts (see, e.g., \cite{FreedRaj}, particularly Figures 1–3, and the references therein). More broadly, the model presented here offers a simple alternative to classical elastic string models where standard constitutive assumptions fail to incorporate natural stretch-limiting behavior.}

\bigskip

\centerline{\scshape R. Bustamante}
\smallskip
{\footnotesize	
	\centerline{Departamento de Ingenier\'{\i}a Mec\'anica, Universidad de Chile}
	
	\centerline{Beauchef 851, Santiago Centro,  Santiago, Chile}
	
	\centerline{\email{rogbusta@ing.uchile.cl}}
}

\bigskip

\centerline{\scshape K. R. Rajagopal}
\smallskip
{\footnotesize
	\centerline{Department of Mechanical Engineering, Texas A\&M University}
	
	\centerline{College Station, TX 77843, USA}
	
	\centerline{\email{krajagopal@tamu.edu}}
}

\bigskip

\centerline{\scshape C. Rodriguez}
\smallskip
{\footnotesize
	\centerline{Department of Mathematics, University of North Carolina}
	
	\centerline{Chapel Hill, NC 27599, USA}
	
	\centerline{\email{crodrig@email.unc.edu}}
}

\end{document}